\documentclass[rnote]{aa}
\usepackage{graphicx}
\usepackage{txfonts}
\usepackage{natbib}
\bibpunct{(}{)}{;}{a}{}{,}
\newcommand{\Halpha}{\ensuremath{\mathrm{H}\alpha}}

\newcommand{\kms}{\ensuremath{\mathrm{km}\,\mathrm{s}^{-1}}}
\newcommand{\kupa}{NGC\,6910}
\newcommand{\wpet}{BD\,+40$^\circ$4146}
\newcommand{\wtri}{BD\,+40$^\circ$4148}

\begin{document}

\title{The H$\alpha$ stellar and interstellar emission in the open
cluster \kupa}

\author{Ji\v{r}\'{\i} Kub\'at\inst{1}
    \and
    Daniela Kor\v{c}\'akov\'a\inst{1}
    \and
    Ad\'ela Kawka\inst{1}
    \and
    Andrzej Pigulski\inst{2}
    \and
    Miroslav \v{S}lechta\inst{1}
    \and
    Petr \v{S}koda\inst{1}}

\offprints{J. Kub\'at}

\institute{Astronomick\'{y} \'{u}stav, Akademie v\v{e}d \v{C}esk\'{e}
    republiky, Fri\v{c}ova 298, CZ-251 65 Ond\v{r}ejov,
    Czech Republic
    \and
    Instytut Astronomiczny Uniwersytetu Wroc{\l}awskiego,
    ul. Kopernika 11, 51-622 Wroc{\l}aw, Poland}

\date{Received 26 January 2007 / Accepted 8 June 2007}

  \abstract
{}
{We verify the nature of emission-line 
stars in the field of the open cluster \kupa.}
{Spectroscopy in the {\Halpha} region was obtained.
Raw CCD frames of spectra of all stars fainter than $V$ = 9\,mag
observed by us are significantly affected by nebular emission
originating in the surrounding \ion{H}{ii} region IC\,1318.
After careful data reduction and subtraction of the nebular 
radiation we succeeded
in obtaining
reliable stellar spectra.
}
{
We
confirm
that the star {\kupa}~37 is a Be star, and we
have
corrected the
classification of V1973 Cyg from an Ae star to a normal A type star.
Since the diffuse interstellar bands do not appear in the spectrum of
this star while being present in the other stars we observed, we confirm
that V1973 Cyg is a foreground object with respect to IC\,1318
and {\kupa}.
We also
find
that the H$\alpha$ line in HD\,194279 has a P~Cygni
profile and the H$\alpha$ line profile is
variable in HD\,229196.
}
{}
   \keywords{
   Open clusters and associations: individual: {\kupa} --
   stars: emission-line, Be --
   stars: individual: V1973\,Cyg --
   ISM: individual objects: IC\,1318
   }

\authorrunning{J. Kub\'at et al.}

   \maketitle
%

\section{Introduction}

The
northern-sky
open cluster {\kupa}, the core of the Cyg\,OB9
association, lies in the vicinity of the nebula IC\,1318b, a part of
IC\,1318 -- the $\gamma$ Cygni Nebula.
\cite{rusky} identified
the star GSC 03156-00657 (indicated as No.~1 in their Fig.~1)
as the excitation source of this nebula.
According to \cite{nemci}, it is an O9\,V type star.
However, this star is not a member of
{\kupa}, because
it lies too far from
the cluster centre.
The IC\,1318b nebula harbours a star-forming region SFR~2\,Cyg,
in which many emission line stars are known \citep{ms90,shevch,manch}.
In the field of the cluster, however, only a few stars with emission
were found.
Considering the same area as \cite{otw6}, i.e., 20$'$ $\times$ 20$'$
centred
at {\kupa}, seven stars with {\Halpha} emission were reported
in the literature.
However, only four of them are brighter than $V$ = 13~mag.
The brightest, HD\,194279, was originally classified as a B0e star
(MWC\,634) by \citet{mwc3}.
Later, \citet{mcw} classified this supergiant as B1.5\,Ia (MCW\,896).
Following the discovery of its light variations by Hipparcos, the star
was named V2118\,Cyg \citep{varnames99}.

\citet{Beopen} did not find any Be stars in this cluster after checking
stars brighter than 11.0 mag.
A negative result was also obtained by \citet{Beyoung}.
On the other hand, \citet{hal91} found weak central emission in the
{\Halpha} line in the spectrum of the A3 star
HD\,229189 = {BD\,+40$^\circ$4145}
\citep[V1973\,Cyg,][]{varnames93}
and classified this star as an Ae star.
Usually only the hottest A-type stars are Ae stars, therefore
this star seemed to be quite exceptional.

Recently, \cite{otw6} searched {\kupa} for variable B type stars.
They also looked for emission-line objects in this cluster using
{\Halpha} photometry.
They
find
three stars that have {\Halpha} in emission.
To the previously known emission line star
V2118 Cyg,
they added an
O-type star HD\,229196 \citep[V2245\,Cyg,][]{varnames01} = MCW\,895
\citep{mcw},
which was
previously known
only from some
peculiarities in the ultraviolet spectrum \citep{massa}, and 
{\kupa}~37.\footnote{See the WEBDA database 
(http://obswww.unige.ch/webda/) for the numbering system used in
{\kupa}.}
This research note aims to
verify these findings spectroscopically.
%

\section{Observations and data reduction}

\begin{figure}[b]
\centering
\resizebox{\hsize}{!}{\includegraphics{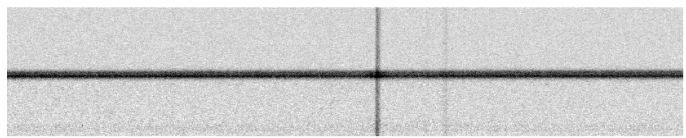}}
\caption{CCD image of the slit spectrum of {\kupa}~37.
Note the vertical lines corresponding to the 
emission from the nebula.
The brighter line corresponds to the {\Halpha} line, the
fainter
to the [\ion{N}{ii}] 6583.45\,{\AA} line.}
\label{widmo}
\end{figure}

\begin{table*}
\caption{List of Ond\v{r}ejov
coud\'e observations of stars in {\kupa}.
Suspected emission objects are listed in the upper part of
the table followed by the comparison stars.
The $V$ magnitudes are from \cite{otw6}.}\label{pozorovani}
\begin{center}
\begin{tabular} {cclrrccr}
\hline
WEBDA & Other star & \multicolumn{1}{c}{Spectral} &
\multicolumn{1}{c}{$V$} & \multicolumn{1}{c}{HJD} &
Spectral & \multicolumn{1}{c}{Exposure} \\
No. & names & \multicolumn{1}{c}{classification}
& \multicolumn{1}{c}{[mag]} & 2450000+ &
region [\AA] & \multicolumn{1}{c}{time [s]} \\
\hline
2 & V2118 Cyg (HD\,194279) & B1.5\,Ia$^2$, B2\,Ia$^4$ &
7.12 & 3985.3520 & 5952 -- 7179 & 300 \\
& & & & 3999.5324 & 5952 -- 7179 & 600 \\
\noalign{\smallskip}
4 & V2245 Cyg (HD\,229196) & O6\,III$^5$ &
8.61 & 3941.5973 & 5952 -- 7179 & 678 \\
& & & & 3985.3606 & 5952 -- 7179 & 300 \\
& & & & 3999.5208 & 5952 -- 7179 & 900 \\
\noalign{\smallskip}
6 & V1973 Cyg (HD\,229189) & A3\,V$^3$, A3\,Ve$^6$ &
10.12 & 3941.5719 & 5952 -- 7179 & 2700 \\
\noalign{\smallskip}
37 & & Be$^7$ & 12.44 & 3939.3843 & 5952 -- 7179 & 2700 \\
 & & (from H$\alpha$ & & 3939.4182 & 5952 -- 7179 & 2700 \\
 & & photometry)& & 3939.4656 & 6160 -- 6772 & 3600 \\
\noalign{\smallskip}
\hline
\noalign{\smallskip}
5 & \wpet & B3$^1$ & 9.74 & 3939.5126 & 6160 -- 6772 & 3600 \\
\noalign{\smallskip}
13 & \wtri & O9.5\,V$^3$ & 10.36 & 3941.5351 & 5952 -- 7179 & 2700 \\
\noalign{\smallskip}
14 & & B0.5\,V$^3$ & 10.46 & 3941.4782 & 5952 -- 7179 & 2700 \\
\noalign{\smallskip}
25 & & B3$^1$ & 11.52 & 3950.4634 & 5952 -- 7179 & 3600 \\
& & & & 3950.5186 & 5952 -- 7179 & 3261 \\
\hline
\end{tabular}
\end{center}
{\em Note:} Spectral types are as classified by:
1 -- \cite{bs49},
2 -- \citet{mcw},
3 -- \cite{ha65},
4 -- \cite{obtype},
5 -- \cite{otype},
6 -- \cite{hal91},
7 -- \cite{otw6}.
\end{table*}

We observed four suspected {\Halpha} emission line stars
brighter than 13 mag in $V$:
V2118\,Cyg, V2245\,Cyg, V1973\,Cyg, and {\kupa}~37.
In addition, for comparison purposes we observed several other
stars that
are brighter than and
near
{\kupa}~37, namely: \wpet, \wtri, {\kupa}~14, and {\kupa}~25.

Our spectroscopic observations were obtained in 2006 using the
spectrograph in the
coud\'e focus of the Ond\v{r}ejov 2-m telescope;
for a description see \citet{ccdpopis}.
We observed the region around the {\Halpha}
line, and the observations are summarised in Table~\ref{pozorovani}.
We obtained medium-resolution spectra ($R\sim$ 13\,000) over the
wavelength region 6160 -- 6672\,{\AA} using the 700-mm spectrograph
focus and low-resolution spectra ($R\sim$ 7400) over the wavelength
region 5952 -- 7179\,{\AA} using the 400-mm spectrograph focus.
In the 700-mm focus we used the SITe CCD 2030 $\times$ 800 chip having
15-$\mu$m pixels, while in the 400-mm focus we used the
LORAL CCD 2688 $\times$ 512 chip
%
with 15-$\mu$m pixels.

The data were reduced using the Image Reduction and Analysis Facility
(IRAF\footnote{IRAF is distributed by the National Optical Astronomy
    Ob\-ser\-va\-to\-ries, which are operated by the Association of
    Universities for Research in Astronomy, Inc., under cooperative
    agreement with the National Science Foundation.}).
The cluster {\kupa} is in the same region of the sky as the $\gamma$
Cygni Nebula, which is also a significant source of {\Halpha} radiation.
Consequently, an {\Halpha} emission line can be found in the image of
the CCD frame on both sides of the 
slit's
stellar spectrum (see Fig.\,\ref{widmo}).
This fact had to be taken into account in the data reduction process, at
least for the faint stars, for which the contribution from the nebula
becomes non-negligible in comparison with that of the star itself.
After the overscan correction, bias subtraction, and flat field
correction in IRAF, we cleaned the 2D FITS images from
cosmics using the program \texttt{dcr} written by \citet{pych}.
Then, the contribution from the background emission was subtracted.
The removal of cosmics mentioned above enabled us to use columns
(perpendicular to the image of the stellar spectrum)
that are
only one pixel wide.
Then, the background value in the given pixel was obtained using a
second-order
Chebyshev polynomial fit.
The remaining steps, such as wavelength
calibration,
were done using 
IRAF.

\section{The emission line stars in \kupa}

\subsection{\kupa~37} \label{websec}

\begin{figure}[ht!]
\centering
\resizebox{0.9\hsize}{!}{\includegraphics{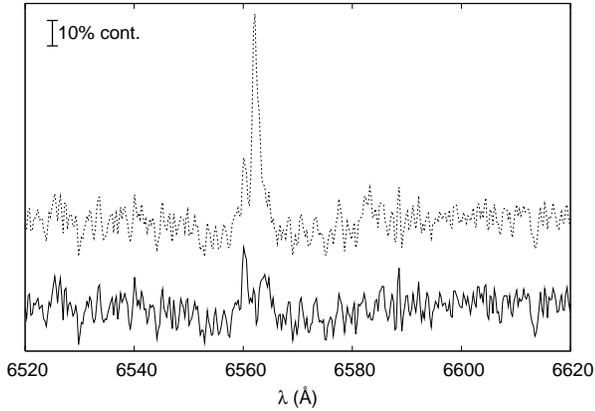}}
\caption{The spectrum of {\kupa}~37 before (dotted line) and after
(continuous line) subtraction of the contribution from the nebula.
Note the presence of the two emission peaks in the first spectrum. The
stronger peak corresponds to the {\Halpha} line, the weaker, to the
nitrogen forbidden line [\ion{N}{ii}] 6583.45\,{\AA}.
The spectra are
normalised
to the continuum and 
separated arbitrarily for clarity.}
\label{w37-700}
\end{figure}

\citet{otw6}
report that {\kupa}~37 is a new Be star,
based on
their {\Halpha} photometry observations.
However, the spectrogram in Fig.~\ref{widmo} shows the presence of
strong background {\Halpha} emission, evidently originating from the
surrounding nebula.
Figure\,\ref{w37-700} shows that,
after background subtraction, emission in
H$\alpha$ is reduced albeit still present.
A similar result was obtained using the
low-dispersion
spectrograms of
this star (not shown here).
This residual emission has to be intrinsic to \kupa~37 and indicates
that the classification of this star as a Be star is likely to be
correct.

In order to confirm the origin of the background emission, we observed
four nearby
stars:
the star {\wpet} = {\kupa}~5, which lies very
close to \kupa~37, the suspected binary {\wtri} = \kupa~13, and two
recently discovered $\beta$~Cephei-type variables {\kupa}~14 and
{\kupa}~25 \citep[see][]{viden}.
All these stars show a central emission peak in their spectra
prior to background subtraction.
The peak disappears once the background contribution is removed.

Since the procedure of background subtraction completely removes the
H$\alpha$ emission from the spectra of \wpet, \wtri, \kupa~14, and
\kupa~25, it is clear that they are not Be stars.
The emission in their uncorrected spectra comes from the surrounding 
nebula. 

\subsection{Bright stars from \kupa}

To
complete the survey of the known emission line stars in
{\kupa}, we observed the brightest stars in the field of the cluster.
Some of these were indicated
to be in emission by \citet{otw6}.
Moreover, they are often used for interstellar absorption studies
\citep[e.g.][and references
therein]{interpendl,interjap,uzkediby,diby,aroma,redd01,redd03,dibrev}.

\subsubsection{V2118 Cyg}

\begin{figure}[ht!]
\centering
\resizebox{0.9\hsize}{!}{\includegraphics{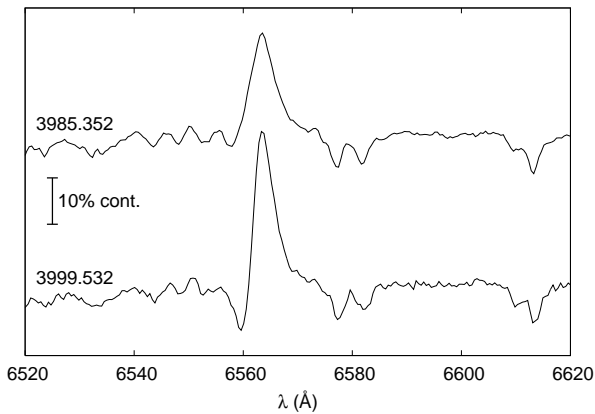}}
\caption{Two spectra of the {\Halpha} line of V2118 Cyg.
The spectra are
labelled
with the HJD$-2450000$.}
\label{v2118}
\end{figure}

The B1.5\,Ia (or B2\,Ia) supergiant HD\,194279 (V2118\,Cyg = \kupa~2)
occasionally appears in radiatively driven wind studies.
It shows P~Cygni profiles for lower hydrogen lines (see
Fig.\,\ref{v2118} for {\Halpha} and Fig.\,4 in \citealp{annique} for
Br$\alpha$).
However, other hydrogen infrared lines and also higher Balmer lines
(H$\beta$ to H$\varepsilon$) were found to be in absorption
\citep{danny92}.

Due to its brightness, the influence of the interstellar emission on the
stellar spectrum of V2118~Cyg is negligible.
Our observations indicate some variability in the P Cygni-type {\Halpha}
line profile (Fig.\,\ref{v2118}).

\subsubsection{V2245 Cyg}

\begin{figure}[t]
\centering
\resizebox{0.9\hsize}{!}{\includegraphics{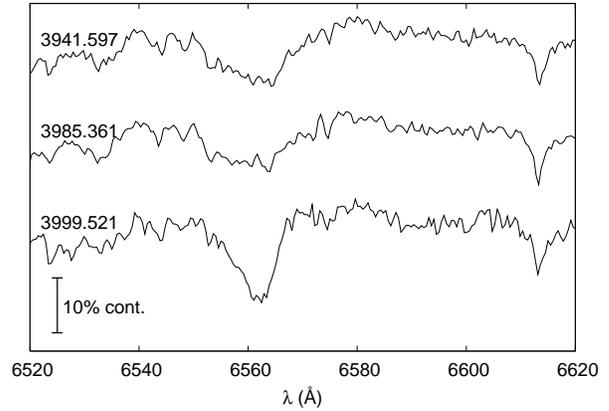}}
\caption{Three spectra of the {\Halpha} line of V2245 Cyg.
The spectra are
labelled
with the HJD$-2450000$.}
\label{v2245}
\end{figure}

The O-type star HD\,229196 (V2245 Cyg = \kupa~4)
was listed
as
possibly in emission by \cite{otw6}.
Apart from being listed in catalogues of Galactic O-type stars
\citep[e.g.][]{okatalog}, this star has not been
studied in detail.
We see no noticeable influence of the background interstellar emission
on the stellar spectrum of V2245\,Cyg, apparently due to its brightness.
Moreover, there is no emission present in any of the three
spectra
(Fig.\,\ref{v2245}); however,
changes in the {\Halpha} line profile are clearly seen.
These changes and the presence of the extended blue wing in the
{\Halpha}
profile,
together with the photometric variability
make this star an interesting object for further study.

\subsection{V1973 Cyg}

The observations of \kupa~37 and nearby stars discussed in
Sect.~\ref{websec} lead us to
suspect
that the central {\Halpha}
emission of V1973 Cyg (\kupa~6), which was reported by \cite{hal91}, may
also
have an
interstellar origin.
The result for this star was that again the central emission disappeared
after the background was subtracted.
Since without the subtraction of the background we obtained 
spectra similar to that presented by \cite{hal91}, we suspect that 
the spectrum she showed in
this
paper might be affected by the
nebular emission. 
Thus, V1973 Cyg probably does not belong to the class of Ae stars.
However, due to its peculiar variability found by \citet{v1973var},
and
confirmed by \citet{hal91} and recently by \citet{viden}, this star
remains a very interesting object for further study.

\begin{figure}[ht!]
\centering
\resizebox{0.9\hsize}{!}{\includegraphics{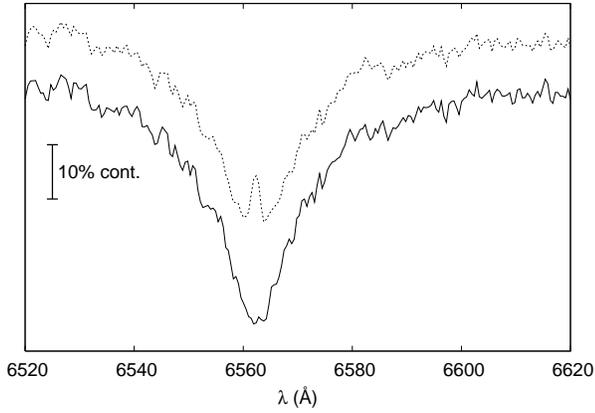}}
\caption{The same as in Fig.\,\ref{w37-700}, but for V1973 Cyg.}
\label{v1973}
\end{figure}

\section{Radial velocities and cluster membership}

\begin{table}[b]
\caption{Heliocentric radial velocities of the observed stars
($RV_\mathrm{star}$) and surrounding nebula ($RV_\mathrm{neb}$) in
{\kupa} as measured using the {\Halpha} line.}
\label{rvel}
\begin{tabular}{ccc}
\hline
Star & $RV_\mathrm{star}$ [\kms] & $RV_\mathrm{neb}$ [\kms] \\
\hline
\wpet     & $-$17.8 & $-$28.3 \\
V1973 Cyg & $-$6.9 & $-$24.7 \\
\wtri & $+$11.4 & $-$26.0 \\
\kupa~14 & $-$20.1 & $-$29.7 \\
\kupa~25 & $-$7.3; $-$8.2 & $-$26,5; $-$25.1 \\
\kupa~37 &
$-$54.8; $-$46.5; $-$32.4 & $-$27.9; $-$27.0; $-$26.0 \\
\hline
\end{tabular}
\end{table}

\begin{figure}[h]
\centering
\resizebox{0.43\hsize}{!}{\includegraphics{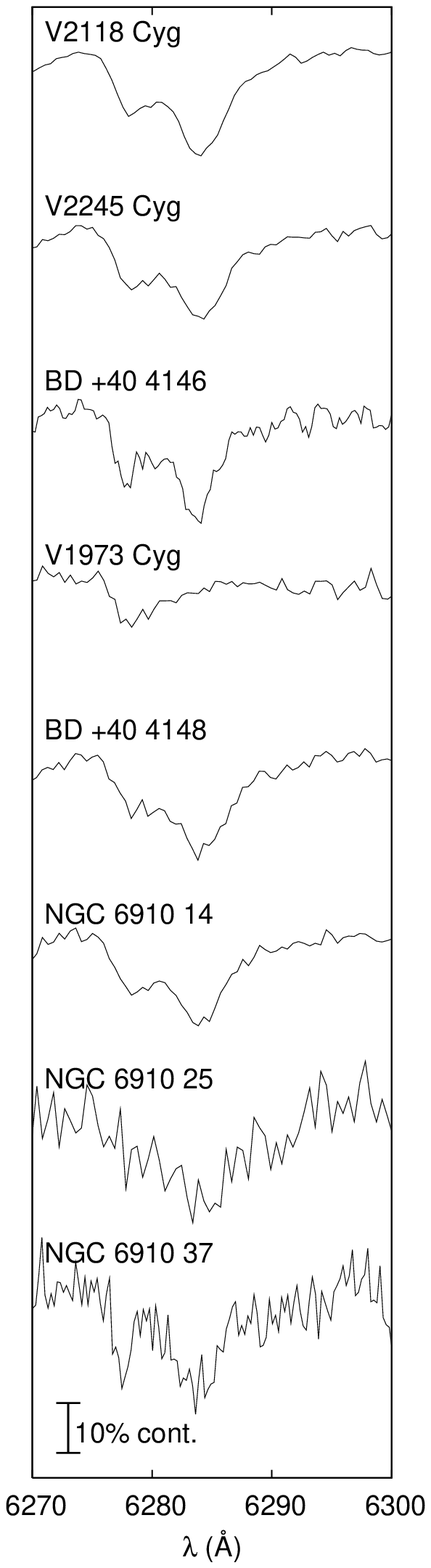}}
\resizebox{0.43\hsize}{!}{\includegraphics{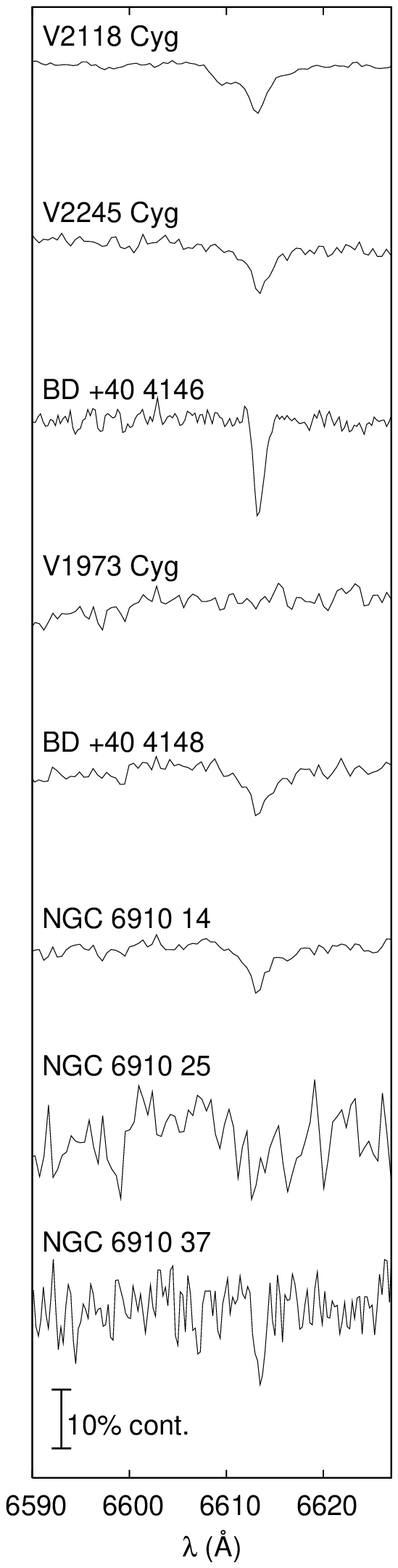}}
\caption{Diffuse interstellar bands at 6284{\AA} and 6614{\AA} in the
spectra of the observed stars.
The absorption next to 6284\,{\AA} DIB, at about 6278\,{\AA}, is a
telluric line.
The spectrum of \kupa~25 is too noisy for the 6614\,{\AA} DIB to be
visible.
The spectra of {\wpet} and \kupa~37 were observed at a higher
resolution; consequently, the DIB at 6614\,{\AA} is deeper and narrower
for these two stars.
Note the absence of both DIBs in the spectrum of V1973\,Cyg.}
\label{diby}
\end{figure}

To clarify their cluster membership, we measured the heliocentric radial
velocities with an accuracy of $\sim2$\,{\kms} of six of the observed
stars in
{\kupa},
as well as the heliocentric radial velocity of the
IC\,1318 nebula in the vicinity of each observed star (see
Table~\ref{rvel}).
All measured stars except V1973\,Cyg are cluster members.
Cluster membership of V1973\,Cyg
has already been
ruled out by \cite{bs49}.
According to \cite{ha65}, it is a foreground
object, and
\cite{shevch} came
to the same conclusion.
Out of the five measured cluster 
members,
only one is non-variable, while
the remaining are either pulsating stars (\kupa~14 and \kupa~25) or
suspected binaries \citep[\wtri,][]{otw6}.
Since only one or two spectra were obtained for each star, then the
measured radial velocities are not representative of the
centre-of-mass
velocities.
In consequence, we can only conclude that the radial velocities we
measured
do not contradict
the cluster membership as
proposed by the photometric data.
The radial velocities of {\kupa}~37 display significant changes during
the three hours of its observation (see
Table~\ref{pozorovani})
and
suggest possible binarity of this Be star.

The measured radial velocities of the nebular H$\alpha$ line
(Table~\ref{rvel})
are between $-$30 and $-$25\,$\kms$, which is
consistent with the results of \citet[][see their Fig.\,5]{ic1318b}.
Radio observations by \citet{dato63} found that an expanding shell of
neutral hydrogen with a radial velocity of about $-$28.5\,{\kms}
surrounds the cluster and the \ion{H}{ii} region.
The radial velocity of the cluster is $-$31\,{\kms} \citep{hayf32},
which is similar to the velocity of the neutral hydrogen shell.
Also the radial velocity of the emission peak in the {\Halpha} line
of V1973\,Cyg, published by \cite{hal91}, is $-29.6\pm2.2\,{\kms}$,
which supports the fact that the central emission in the {\Halpha}
profile of this star
has an
interstellar origin.

Support for the cluster membership of stars comes from the presence of
diffuse interstellar bands (DIBs) in the stellar spectra.
DIBs at 6614 and 6284\,{\AA} \citep{uzkediby,diby} are observed in all
stars except V1973 Cyg.
The absence of the two DIBs in the spectrum of the latter star and its
presence in the spectra of the other stars proves that V1973\,Cyg is a
foreground object that is much closer to us than {\kupa}.
The presence of a DIB does not prove that the remaining seven stars are
members of {\kupa}, but the similarity of their DIB profiles
(once resolution is taken into account) in 
Fig.~\ref{diby} suggests that these stars have very similar distances.
If we adopt the measured values of the DIB wavelengths from
\cite{uzkediby}, 6613.56 and 6283.85\,\AA, their radial velocities have
values around $-$9\,$\kms$.
However, since the nature of these DIBs is not known, the exact
wavelengths are uncertain and, consequently, measured values of radial
velocities may differ systematically from those of the nebular
$\Halpha$ lines.

The cluster itself becomes very interesting in view of the fact that it 
harbours at least seven $\beta$~Cephei stars
\citep{viden},
which makes
it one of the richest in this type of variable.
Because the $\beta$~Cephei stars constitute a relatively large fraction
of early B-type cluster members, it was suggested
\citep{otw6} that this can be explained by higher metallicity.
The pulsations and Be phenomenon are not mutually exclusive,
but the pulsations are enhanced in the high metallicity environment,
while
the effect is opposite for the Be phenomenon
\citep{wibj06,mart06}.
Moreover, the fraction of Be stars may depend on the age of a cluster.
Because of the small number statistics, no reliable conclusion as to the
observational relation between the incidence of $\beta$~Cephei and Be
stars can be drawn
%
at the moment.

\section{Conclusions}
Our spectroscopic observations of a sample of eight stars in {\kupa}
have
confirmed the weak emission in the star
\kupa\,37 and,
consequently,
its classification as a Be star. There is, however, a significant
contribution from the surrounding nebula to the observed spectra, which
needs to be subtracted for stars observed in this region of the sky.
We argue that this is probably the reason
%
the variable star
V1973\,Cyg was thought to be an Ae
star,
while our observations show that
it is a normal A-type star. 
The radiation from the nebula needs to be considered when observing
stars in this region of the sky. Otherwise, as we have shown in the
paper and, for example, as shown by \citet{kebe98} for NGC~330, spurious
emission can be detected.

We also confirm the weak emission in the supergiant HD\,194279 and find
P Cygni-type profiles of its {\Halpha} line.
There is no emission in the H$\alpha$ line of the other bright star in
the cluster, HD\,229196, but its {\Halpha} profile is variable.
It is therefore possible that at the epoch of the photometric
observations, the {\Halpha} profile was partly filled in by emission
that
would be sufficient to get an agreement with the $\alpha$ index
measured by \citet{otw6}.

\begin{acknowledgements}
This research
%
made use of the NASA Astrophysics Data System (ADS)
and of the SIMBAD database.
The authors appreciate the technical support for the introduction of the
400-mm focus camera by F.\,\v{Z}\v{d}\'arsk\'y, J.\,Honsa, and
J.\,Fuchs.
The authors would like to thank the anonymous referee for valuable
comments.
%
This research was supported by grants GA AV \v{C}R B301630501
and GA \v{C}R 205/06/0584.
The Astronomical Institute Ond\v{r}ejov is supported by a project
AV0\,Z10030501.
\end{acknowledgements}

\end{document}